\documentclass[prl,twocolumn,amsmath,amssymb,floatfix,superscriptaddress]{revtex4-1}
\usepackage{color,graphics,epsfig,rotating}
\usepackage{graphicx}% Include figure files
\usepackage{dcolumn}% Align table columns on decimal point
\usepackage{bm}% bold math
\usepackage{mathrsfs}
\usepackage{lmodern}
\usepackage{notoccite}

\usepackage{grffile}
\usepackage{xspace}
\usepackage{xifthen}
\usepackage{soul}

\graphicspath{{figs/}}

\usepackage[svgnames]{xcolor}
\newif\ifshowcomments\showcommentsfalse

\newcommand{\ce}[1]{\ifshowcomments{\color{Blue}[\textbf{CE: }\textit{#1}]\color{black}}\else{}\fi}

\usepackage[final]{changes}

\def\beq{\begin{equation}}
\def\eeq{\end{equation}}

\newcommand{\rno}[1]{{#1}NiO$_{3}$}

\newcommand{\eg}{e_{g}}

\newcommand{\xx}{{x^{2} - y^{2}}}
\newcommand{\zz}{{z^{2}}}

\newcommand{\ndisp}{\nu}
\newcommand{\bdisp}{\Delta^{s}}

\newcommand{\pd}{\partial}

\newcommand{\Hint}{H_{\mathrm{int}}}
\newcommand{\Hband}{H_{\mathrm{band}}}
\newcommand{\Hellatt}{H_{\mathrm{e-l}}}
\newcommand{\Hlatt}{H_{\mathrm{latt}}}

\newcommand{\LB}{\textsc{lb}}
\newcommand{\SB}{\textsc{sb}}

\newcommand{\glat}{g}
\newcommand{\eps}{\varepsilon}
\newcommand{\Av}[1]{\left\langle{#1}\right\rangle}

\begin{document}

\title{Mechanism and Control Parameters of the Coupled Structural and Metal-Insulator Transition in
Nickelates
}

\author{Oleg E. Peil}
\affiliation{Materials Center Leoben Forschung GmbH, Roseggerstra{\ss}e 12, A-8700 Leoben, Austria}
\affiliation{DQMP, Universit\'e de Gen\`eve, 24 quai Ernest Ansermet, CH-1211 Gen\`eve, Suisse}
\author{Alexander Hampel}
%\email{alexander.hampel@mat.ethz.ch}
\affiliation{Materials Theory, ETH Z\"u{}rich, Wolfgang-Pauli-Strasse 27, 8093 Z\"u{}rich, Switzerland}
\author{Claude Ederer}
%\email{claude.ederer@mat.ethz.ch}
\affiliation{Materials Theory, ETH Z\"u{}rich, Wolfgang-Pauli-Strasse 27, 8093 Z\"u{}rich, Switzerland}
\author{Antoine Georges}
\affiliation{DQMP, Universit\'e de Gen\`eve, 24 quai Ernest Ansermet, CH-1211 Gen\`eve, Suisse}
\affiliation{Coll\`ege de France, 11 place Marcelin Berthelot, 75005 Paris, France}
\affiliation{Center for Computational Quantum Physics, Flatiron Institute,
162 Fifth avenue, New York, NY 10010, USA}
\affiliation{CPHT, Ecole Polytechnique, CNRS, Universit\'e Paris-Saclay, 91128 Palaiseau, France}

\date{\today}
%\pacs{71.30.+h,71.15.Mb,71.38.-k}

%
% *. We reveal in details a mechanism of the metal-insulator transition (MIT) coupled with
%    a structural transition associated with the breathing mode (BM).
%
% *. We use both a simplified model and a combination of DFT and DMFT to show that...
%
% *. Proximity to the electronic instability softens the BM.
%
% *. The system is pushed through the transition by the MIT induced by the BD.
%
% *. ...
%
\begin{abstract}
Rare-earth nickelates exhibit a remarkable metal-insulator transition accompanied
by a symmetry-lowering lattice \ce{or ``structural''} distortion.
Using model considerations and first-principles calculations, we present a theory of this phase transition
which reveals the key role of the coupling between electronic and lattice instabilities.
We show that the transition is driven by the proximity to an
instability toward electronic disproportionation which couples to a specific structural distortion mode,
cooperatively driving the system into the insulating state.
This allows us to identify two key control parameters of the transition:
the susceptibility to electronic disproportionation and the stiffness of the lattice mode.
We show that our findings can be rationalized in terms of a Landau theory involving
two coupled order parameters, with general implications for transition-metal oxides.
\end{abstract}

\maketitle

%%%%%%%%%%%%%%%%%%%%%%%%%%%%%%%%%%%%%%%%%%%%%%%%%%%%%%%%%%%%%%%%%%%%%%%%%%%%%%%%
%
% Introduction
%
%%%%%%%%%%%%%%%%%%%%%%%%%%%%%%%%%%%%%%%%%%%%%%%%%%%%%%%%%%%%%%%%%%%%%%%%%%%%%%%%
%Key questions:
%
%\begin{itemize}
%    \item What is the mechanism responsible for the MIT in RNiO3 ?
%    \item Interplay of electronic and structural degrees of freedom: which one is the driving force ?
%    \item What are the control parameters (accounting e.g. for the changes along the RE series) ?
%%   What are the nature and the control parameters of the MI and structural transition in RNiO3?
%\end{itemize}

%\textit{Introduction.}
The coupling of electrons to lattice degrees of freedom %\added{(DoF)}
provides a key opportunity to control the properties of strongly correlated materials, %via the lattice,
as in, e.g., epitaxial heterostructures \cite{Zubko2011}.
%
% <OEP>: mention other oxides?
Such a coupling often leads to concomitant electronic and
structural transitions, which have been observed in V$_{2}$O$_{3}$ \cite{McWhan1969}, manganates \cite{Tokura2006},
Ca$_{2}$RuO$_{4}$~\cite{Nakatsuji1997,han_millis_2018}, etc.
Rare-earth nickelates (\rno{$R$}) \cite{Torrance1992,Medarde1997,Catalan2008}
represent an ideal playground in this respect, because
%<OEP>
their metal-insulator transition (MIT),
tightly associated with a lattice mode,
is easily tunable~\cite{Catalano2018,Middey2016}.
%their metal-insulator transition (MIT)
%is tightly associated with a lattice degree of freedom, often referred to as
%a ``breathing mode'' (BM) \cite{Rondinelli2011}.
%
% <OEP>: relocated to the end of the "punch-line" paragraph
%Experimentally, the MIT in \rno{$R$} can be tuned by the choice of the $R$ cation \cite{Torrance1992},
%by epitaxial strain or by more sophisticated methods \cite{Catalano2018,Middey2016}.

%
% What is consensually understood about the transition in RNO?
%
% * Insulating phase is associated with BD (metallic is symmetric)
%
% * Insulating phase is characterized by the negative CT energy with d8L2-d8 configurations
%   formed inside SB-LB octahedra (Demourgues1993,Mizokawa2000,Mandal2017,Millis)
%
% * Low-energy picture of this configuration: eg0-eg2 configuration and U - 3 J < 0 (Subedi,Seth)
%
% * Insulating phase with BD has the lower energy for RNO exhibiting the MIT (Park/Millis)
%
% <OEP>
The MIT in \rno{$R$} is accompanied by a bond disproportionation (BD),
i.e., a coherent contraction of the NiO$_6$ octahedra on one
sublattice [short-bond (SB) octahedra] and an expansion of the
octahedra on the other sublattice [long-bond (LB) octahedra], also
referred to as the ``breathing mode''
(BM)~\cite{Medarde1997,Rondinelli2011}.  The resulting ``bond-disproportionated
insulator'' (BDI) is also characterized by an electronic disproportionation
(ED), whereby the local configuration of SB octahedra is close to
to $d^{8}\underline{L}^{2}$ and that of LB ones to $d^{8}$~\cite{Demourgues1993,Mizokawa2000,Johnston2014,Mandal2017},
or in terms of ``frontier'' $e_{g}$ orbitals to $e_{g}^{0}$ and $e_{g}^{2}$, respectively~\cite{Mazin2007,Subedi2015,Seth2017,Varignon2017}.
The electron localization on the LB sublattice is the result of a
``site-selective Mott transition'' \cite{Park2012}, occurring
irrespective of the (ground state) magnetic ordering for all systems
with $R$-cations smaller than Nd, and lowering the energy of the insulating phase below that of the metallic phase
\cite{Park2014,Johnston2014,hampel_ederer:2018}. Thereby, magnetic order
seems to play only a secondary role, enhancing an already existing
tendency towards the MIT \cite{Ruppen2017,Haule2017,Post2018}. \ce{Do we need the last sentence? Last half-sentence before that is also a bit unclear.}

%
% What remains to be unclear?
%
% * Why BM is destabilized (driving force of BD)?
%
% * What are the key parameters which determine whether BM is unstable or not?
%
% * Why the transition seems to be always of first-order type, even in the PM phase?
%
% <OEP>
The mechanism of the interplay between electronic and
lattice degrees of freedom is not yet understood. This question is of
key importance to identify the driving force responsible for the BD and for
the first-order transition \cite{Alonso:2001bs} into the paramagnetic (PM) insulating state.
This transition was previously described either as a pure charge-order transition~\cite{Hansmann2017}
or as a result of the coupling between lattice modes only~\cite{Littlewood2017}.
Recently, the authors of Ref.~\cite{Mercy2017} proposed that the transition corresponds to
the gradual softening of the BM mode, associated with the opening of a Peierls gap at the Fermi level, where they used density functional theory (DFT) calculations including the $+U$ correction.
However, this theory describes the transition as second order, in contradiction with experimental
observations~\cite{Alonso:2001bs,Girardot2008}, and furthermore cannot describe
the MIT into the PM state, since DFT+$U$ requires a magnetically ordered state to produce
an insulating gap~\cite{Prosandeev2012,hampel:2017}. Crucially, in the absence of magnetic
ordering, the Peierls gap does not open at the Fermi level~\cite{Subedi2015,Ruppen2015} and
thus cannot be responsible for the insulating nature of the PM phase.

%
% What do we offer?
%
% * Ab-initio based picture of the electronic and lattice aspects of the MIT
%
% * MIT is driven by the proximity to a spontaneous ED, irrespective of the magnetic order.
%
% * MIT induces a strongly non-linear response to the BM, which results in
%   the transition of first order
%
% * We identify key parameters, K, \chi, controlling the transition, which is important
%   for materials design
%
% <OEP>
Here, we present a theory describing the interplay between the electronic and structural aspects
of the MIT. We show that the MIT is driven by the proximity to a spontaneous ED, which
leads to a strongly non-linear electronic response with respect to variations of the BM amplitude,
resulting in a
first-order phase transition. Our theory also identifies the BM stiffness and the electronic
susceptibility at $\mathbf{q} = (\frac{1}{2},\frac{1}{2},\frac{1}{2})$ as key parameters
controlling the transition.
Experimentally, these parameters can be tuned by the choice of the $R$ cation \cite{Torrance1992},
or by epitaxial strain in thin films and heterostructures~\cite{Catalano2018,Middey2016,Liao2018,Georgescu2018}.
We validate our theory by performing combined DFT and dynamical mean-field theory (DMFT)
\cite{georges_dynamical_1996,Kotliar2006} calculations, allowing us to explore the trends across the rare-earth series.
We also rationalize the overall physical picture in terms of a Landau theory
involving two coupled order parameters: the ED and
an order parameter associated with the metallicity of the system.

%%%%%%%%%%%%%%%%%%%%%%%%%%%%%%%%%%%%%%%%%%%%%%%%%%%%%%%%%%%%%%%%%%%%%%%%%%%%%%%%
%
% Results: Model
%
%%%%%%%%%%%%%%%%%%%%%%%%%%%%%%%%%%%%%%%%%%%%%%%%%%%%%%%%%%%%%%%%%%%%%%%%%%%%%%%%
\textit{Model description.}
We start by constructing a simplified model which reproduces the main features of the MIT in \rno{$R$}.
The model retains only the key low-energy degrees of freedom:
the interacting electrons in the two frontier $e_g$ orbitals and the BM amplitude $Q$.
The purely electronic part of the Hamiltonian, $\Hband+\Hint$, consists of a simplified tight-binding
(TB) model:
$\Hband = -\sum_{i, j, m, m', \sigma} t^{m m'}_{ij} d^\dagger_{i m \sigma} d_{j m \sigma}\,$ and a local interaction term $\Hint$.
Here, $i,j$ indicate sites within a simple cubic lattice, $m=1,2$ correspond
to the $d_{\xx}$, $d_{\zz}$ orbitals on each site, and hopping matrices $t^{mm'}_{ij}$
are obtained using the Slater-Koster construction with two hopping amplitudes, $t$ and $t'$,
limited to nearest-neighbor and next-nearest neighbor sites, respectively~\cite{Lee2011,*Lee2011b}.
The interaction term $\Hint$ involves two coupling constants: a repulsive interaction $U$ and
an intra-atomic (Hund's) exchange $J$, and takes the standard 2-orbital Hubbard-Kanamori form \cite{Georges2013}.
The purely lattice part is described by an elastic term:
$\Hlatt = \frac{K}{2} Q^{2}$, with $K$ being the stiffness of the BM.
Finally, and importantly, the coupling of the BM amplitude to the electrons is captured by a term:
$\Hellatt = \frac{1}{2}\sum_{m\sigma} \bdisp_m[Q] \left[
\sum_{i\in\SB} \hat{n}_{im\sigma} - \sum_{i\in\LB} \hat{n}_{im\sigma}
\right]$,
where $\hat{n}_{im\sigma}=d^\dagger_{im\sigma}d_{im\sigma}$ is the electron occupation operator
and $\bdisp_m[Q]$ is a (Peierls-like) modulation of the on-site potential seen by orbital $m$ due to the BM structural
distortion parametrized by $Q$. It couples to the operator measuring the ED between
the LB and SB octahedra.
The total Hamiltonian thus reads: $H =  \Hband + \Hint + \Hellatt + \Hlatt$.

At this stage, we define the amplitude $Q$ as the disproportionation in octahedral bond lengths
$b = b_{0} + Q / 2$ for LB and $b = b_{0} - Q / 2$ for SB octahedra.
The modulation of the on-site potential,
$\bdisp_{m}[Q]$, is \added{given by} the difference between the on-site energies of the SB and LB sites:
$\bdisp_{m}[Q] = \eps_{m}[b_{0} - Q/2] - \eps_{m}[b_{0} + Q/2] \approx
(d \eps_{m} / d b)_{b_{0}} Q \equiv \glat_{m} Q$, where we have expanded in $Q$ and
introduced the electron-lattice coupling parameter $g_{m}$.
%%%%%%%%%%%%%%%%%%%%%%%%%%%%%%%%%%%%%%%%%%%%%%%%%%%%%%%%%%%%%%%%%%%%%%%%%%%%%%%%
%
% Figure 1
%
\begin{figure}
    \includegraphics[width=1.0\linewidth]{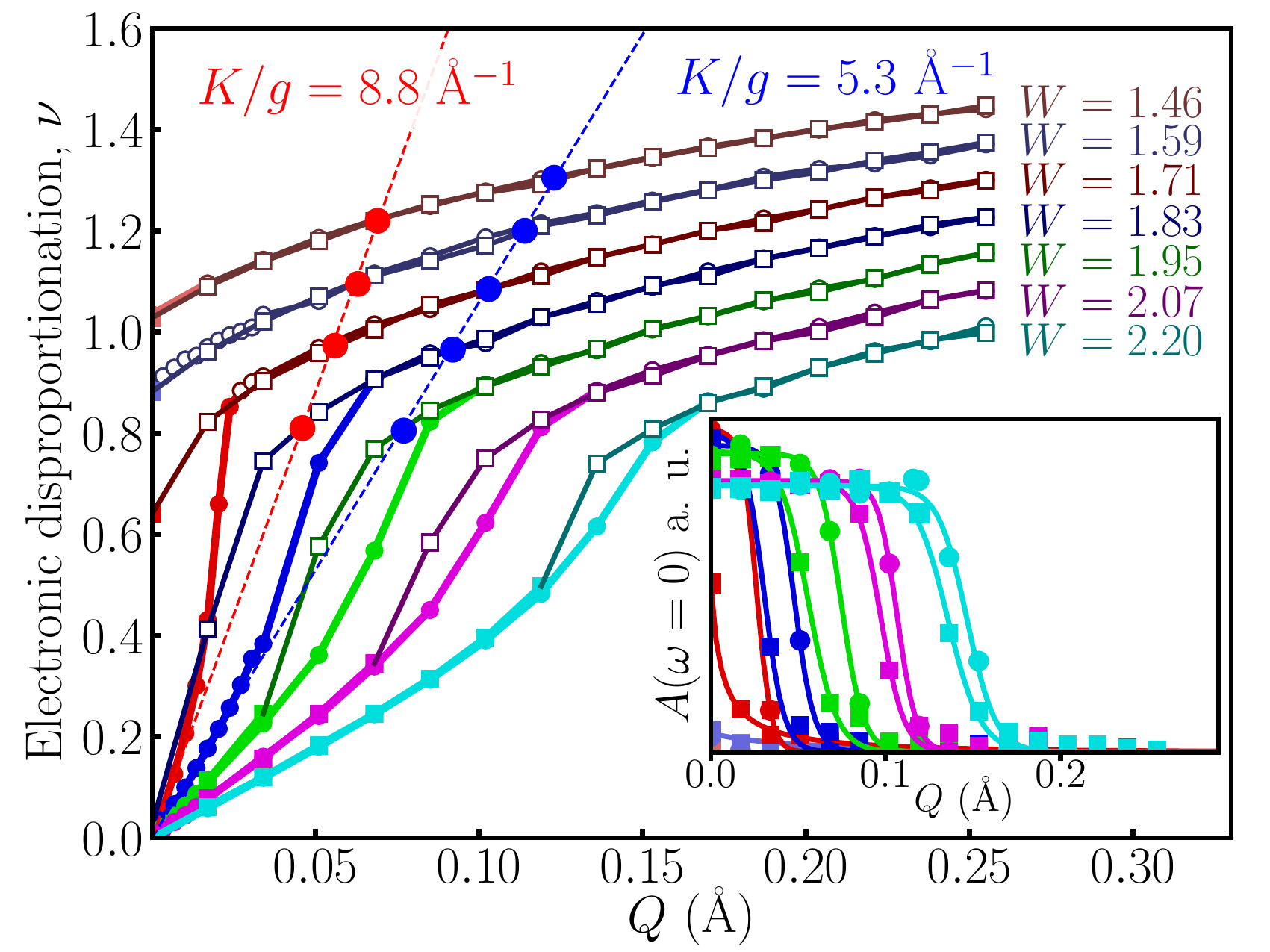}
    \caption{Electronic disproportionation $\nu$ of the TB model as a function of bond disproportionation $Q$ for various values of the bandwidth $W$ (in eV).
Open and filled symbols correspond to insulating and metallic branches, respectively.
The dashed lines represent $(2 K/ g)Q$ for two values of $K/g$, the intersection points
giving the solutions to Eq.~\eqref{eq:q_eq}.
Inset: Spectral weight as a function of $Q$ .}
    \label{fig:bd_vs_q_model}
\end{figure}
%
%%%%%%%%%%%%%%%%%%%%%%%%%%%%%%%%%%%%%%%%%%%%%%%%%%%%%%%%%%%%%%%%%%%%%%%%%%%%%%%%
%
As emphasized in Refs.~\cite{Subedi2015,Seth2017}, an appropriate low-energy description of
the ``negative charge-transfer'' character of \rno{$R$} and of their tendency to form a BDI state
is obtained with $U-3J \lesssim \Delta_s$.
In this regime, the orbital polarization is strongly suppressed
implying that the on-site energies are to a good approximation orbital-independent: $\eps_{\zz} \approx \eps_{\xx}$.
We thus assume that the $\eg$ states are degenerate and omit the index $m$
in one-electron quantities (i.e., $g_m=g$, $\Delta^{s}_{m} = \Delta_{s}$)
\footnote{We checked that in all our DMFT calculations the orbital occupation numbers always ended
up being nearly equal.}.

% <OEP>: expanded the expression for the energy
Minimizing the total energy,
$E = \Av{H} \equiv E_{\textrm{el}}[\nu] - g Q \nu / 2 + K Q^{2} / 2$, with respect to $Q$ yields:
\begin{equation}
    \frac{2K}{g}\,\bar{Q}\,=\,\ndisp[\bar{Q}],
\label{eq:q_eq}
\end{equation}
where $\ndisp[Q] = \Av{\hat{n}_{\LB} - \hat{n}_{\SB}}_{Q}$ is the average ED for a given amplitude $Q$.
Eq.~\eqref{eq:q_eq} is the central equation of this article: It enables one to determine
the equilibrium BM amplitude from the knowledge of the electronic response encoded in
$\ndisp[Q]$ for a given lattice stiffness, $K$, and electron-lattice coupling, $g$.
The stability of the solutions of this equation is determined by the renormalized stiffness
$\kappa \equiv\frac{\pd^{2} E}{\pd Q^{2}} = K - \frac{g^{2}}{2} \frac{\pd \ndisp}{\pd \bdisp}$.
Assuming no spontaneous ED ($\ndisp[Q = 0] = 0$), we obtain
$\kappa = K - \chi_{e} g^{2} / 2$, where $\chi_{e} \equiv (\pd \nu / \pd \bdisp)_{Q = 0}$ is
the electronic susceptibility associated with a ``charge'' modulation at a wave-vector
$\mathbf{q} = (\frac{1}{2}, \frac{1}{2}, \frac{1}{2})$ \cite{suppl}.
Hence, the (linear) stability of the high-symmetry phase is controlled by the electronic
response, $\chi_{e}$, which must be compared to $2K/g^2$.

We investigate the solutions to Eq.~\eqref{eq:q_eq} by performing DMFT calculations
to obtain $\nu[Q]$ for various values of the bandwidth, $W$ (see \cite{suppl} for details).
The results are shown in Fig.~\ref{fig:bd_vs_q_model}.
The calculations were performed first for
increasing values, then for decreasing values of $Q$, which resulted in a hysteresis.
The most important feature of this plot is the strongly non-linear dependence of $\ndisp$ on $Q$.
Solutions of Eq.~\eqref{eq:q_eq} are obtained by intersecting $\ndisp[Q]$
with the straight line $(2 K / g) Q$.

This non-linear shape of the $\ndisp[Q]$-curves plays an important role in determining
the nature of the transition. For a given value of
$K/g$, and depending on $W$, there are either one, two, or three intersection
points (disregarding symmetry-equivalent solutions for $Q < 0$).
Consider $K/g=5.3$ \AA{}$^{-1}$. If $W$ is large ($W \geq 2.07$ eV in Fig.~\ref{fig:bd_vs_q_model})
the straight line intersects the ED curve only at $Q = 0$,
rendering it the only solution to Eq.~\eqref{eq:q_eq}.
As $W$ is decreased, it reaches a value below which there
are three intersection points ($W = 1.95$ eV).
The middle intersection point corresponds to an unstable solution ($\kappa < 0$).
The two remaining stable solutions, $Q = 0$ and
$Q = \bar{Q}[K, W] > 0$ mark a coexistence of two different phases. At even smaller values of $W$
($W \lesssim 1.8$) the solution at $Q = 0$
gets destabilized because $\kappa[Q = 0] < 0$.
We are then left with only one solution $Q = \bar{Q}[K, W] > 0$, telling us that only the BD phase is stable here.
The behavior of the solutions of Eq.~\eqref{eq:q_eq} as a function of $W$
tells us that the BD transition is first-order.
Importantly, this is not directly related to the hysteretic behavior of $\ndisp[Q]$.
Even if there was no such hysteresis, the particular non-linear
dependence of $\ndisp[Q]$ would imply that the BD transition is first-order.
The inset of Fig.~\ref{fig:bd_vs_q_model} displays the spectral weight at the Fermi level
as a function of $Q$. We see that the system undergoes a transition from metallic to insulating behaviour
as $Q$ is increased and that the transition regime corresponds to the strongly
non-linear regime of $\ndisp[Q]$.
Hence, it is the MIT which is responsible for the strong non-linearity of $\ndisp[Q]$.
%%%%%%%%%%%%%%%%%%%%%%%%%%%%%%%%%%%%%%%%%%%%%%%%%%%%%%%%%%%%%%%%%%%%%%%%%%%%%%%%
%
% Figure 2
%
\begin{figure}
    \includegraphics[width=1.0\linewidth]{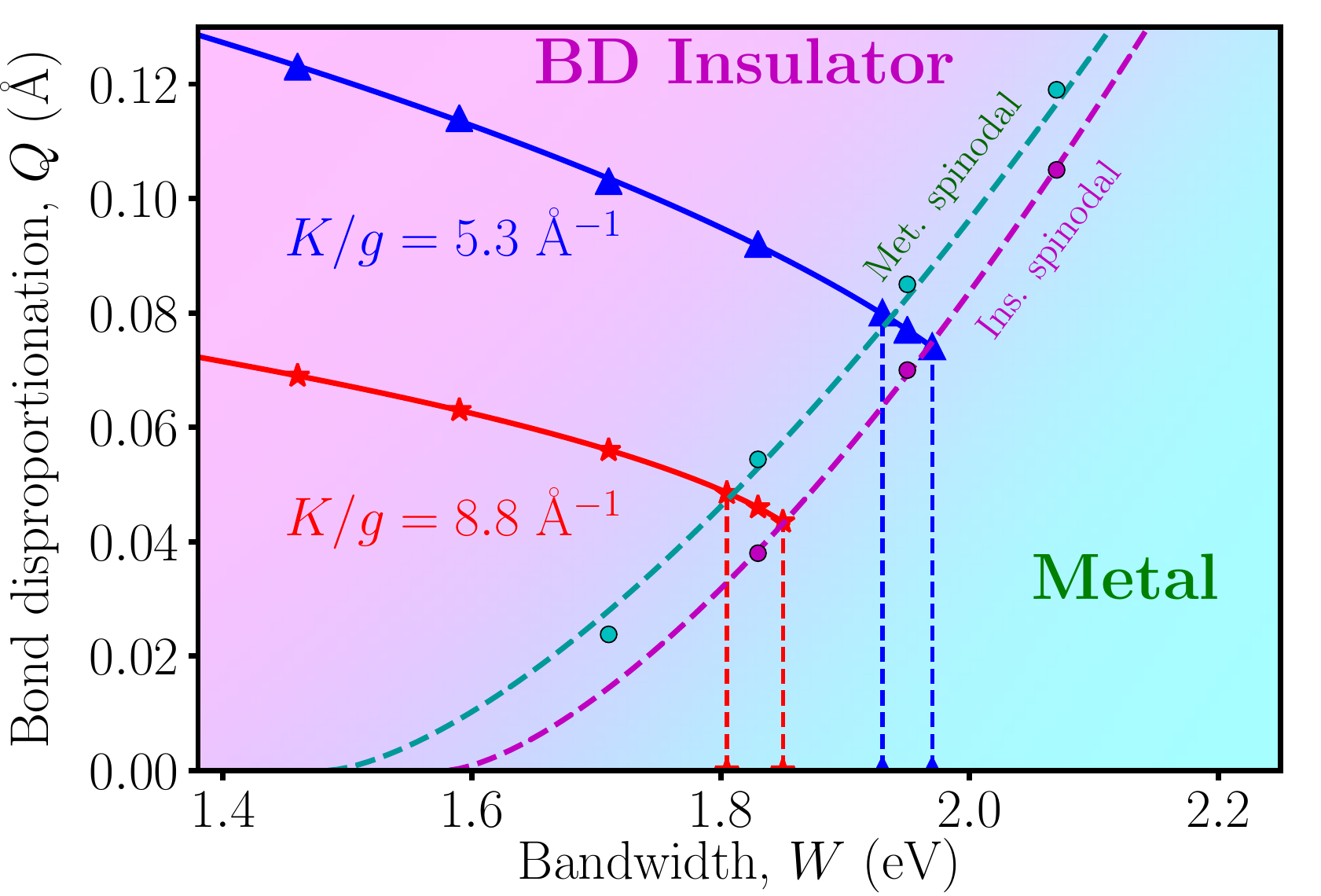}
    \caption{Phase diagram of the TB model as a function of bandwidth $W$ and BD $Q$.
    Circles are critical points for the metallic and insulating phases; dashed curves are spinodal lines obtained
    as fits to   $\alpha (W - W_c)^{\frac{3}{2}}$ (see Landau theory).
    The solid lines with triangles and stars display the stable equilibrium values of $Q$ for two values of $K/g$.
Vertical dashed lines designate lower and upper critical values of $W$ for the structural transition.}
    \label{fig:q_vs_w}
\end{figure}
%
%%%%%%%%%%%%%%%%%%%%%%%%%%%%%%%%%%%%%%%%%%%%%%%%%%%%%%%%%%%%%%%%%%%%%%%%%%%%%%%%

%
% Discussion of control parameters
%
We can now discuss the control parameters of the combined BD/MIT.
Eq.~\eqref{eq:q_eq} shows that the transition behavior
depends on parameters $g$, $K$, and $\chi_{e}$.
%
% <OEP>
%Our DFT calculations reveal that the ratio $K/g$ is almost constant as the rare-earth is
Our DFT calculations reveal that $g$ and $K$ vary little across
the nickelate series (see \cite{suppl}).
Hence, for bulk \rno{$R$}, it is the variation of the electronic susceptibility $\chi_e$ which
plays the key role. The latter is affected by changes in the bandwidth and/or
bond angles~\cite{Hansmann2017}.
In heterostructures and under strain, the BM stiffness, $K$, varies, while $g$ remains unaffected~\cite{Georgescu2018}.
$K$ is thus likely to be an important control parameter in those cases.
This may shed light on the results of Ref.~\cite{Catalano2015} and motivates the variation of $K/g$ in our model calculations.

The effects of the two control parameters $K$ and $\chi_{e}$
(tuned via $K/g$ and $W$, respectively) are summarized in Fig.~\ref{fig:q_vs_w}.
The dashed curves indicate the boundaries (spinodals) of the BDI and
metallic phases, with a narrow coexistence region in-between.
Solid lines show stable non-zero solutions, $\bar{Q}[K, W]$.
For fixed $K$, the bandwidth $W$ (equivalently, $\chi_{e}$)
determines whether the PM ground-state is a BDI phase. %the transition between the two phases being discontinuous.
Variation of $K$ controls smoothly the position of the phase boundaries.

%%%%%%%%%%%%%%%%%%%%%%%%%%%%%%%%%%%%%%%%%%%%%%%%%%%%%%%%%%%%%%%%%%%%%%%%%%%%%%%%
%
% Results: DFT+DMFT
%
%%%%%%%%%%%%%%%%%%%%%%%%%%%%%%%%%%%%%%%%%%%%%%%%%%%%%%%%%%%%%%%%%%%%%%%%%%%%%%%%
\textit{Realistic DFT+DMFT.}
We now perform ab-initio DFT+DMFT calculations to confirm the physics found in
the model calculations and assess materials trends quantitatively.
The impurity model is constructed by projecting onto a low-energy $e_{g}$ subspace
following the scheme described in Refs.~\cite{Subedi2015,Seth2017}
(see \cite{suppl} for details).

%%%%%%%%%%%%%%%%%%%%%%%%%%%%%%%%%%%%%%%%%%%%%%%%%%%%%%%%%%%%%%%%%%%%%%%%%%%%%%%%
%
% Figure 3
%
\begin{figure}
\includegraphics[width=1.0\linewidth]{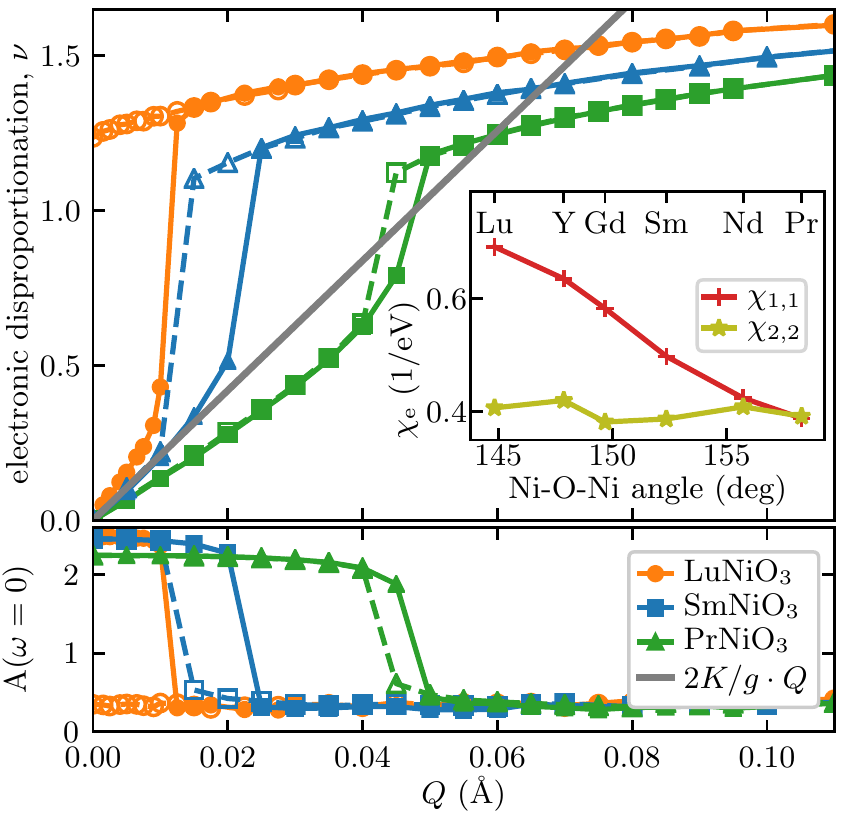}
\caption{Top: ED $\nu$ as a function of
increasing $Q$ (filled symbols) and decreasing $Q$ (open symbols) for $R$=Lu (red), Sm (blue) and Pr (green).
The gray line represents $(2K/g)Q$ with values $g=3.8$~eV/\r{A} and $K=39.3$~eV/\r{A}$^2$ extracted from DFT.
Inset: Electronic susceptibilities as a function of the Ni-O-Ni angle: $\chi_{\text{1,1}}$ is the $d_{x^2-y^2}$ component and $\chi_{\text{2,2}}$ the $d_{z^2}$ component.
Bottom: spectral weight at the Fermi level as a function of $Q$.
}
\label{fig:dft_dmft}
\end{figure}
%
%%%%%%%%%%%%%%%%%%%%%%%%%%%%%%%%%%%%%%%%%%%%%%%%%%%%%%%%%%%%%%%%%%%%%%%%%%%%%%%%

Fig.~\ref{fig:dft_dmft} shows the calculated $\nu(Q)$ for $R$=Lu, Sm, and Pr (top).
The overall nonlinear behavior of $\nu(Q)$ is very similar to that in the model calculations (Fig.~\ref{fig:bd_vs_q_model}),
with the non-linearity clearly related to the MIT (Fig.~\ref{fig:dft_dmft}-bottom),
also indicating a first-order character of the BD/MIT.
This confirms that the model indeed incorporates the essential underlying physics.
Furthermore, we obtain a strong decrease of the amplitude of the non-linearity in $\nu(Q)$ from $R$=Lu towards $R$=Pr,
consistent with the bandwidth variation in the model.
The $d_{x^2-y^2}$ component of the electronic susceptibility
displayed in Fig.~\ref{fig:dft_dmft} (top, inset) increases steadily as the Ni-O-Ni bond angle is reduced (i.e., for for smaller $R$ ions).
This shows that the stronger octahedral rotations for $R$=Lu compared to $R$=Pr lead to %a reduced bandwidth $W$
an increased $\chi_e$~\cite{Lee2011,*Lee2011b,Hansmann2017} and thus control directly the electronic instability.

Finally, the values obtained for $K$ and $g$ from DFT lead to stable equilibrium BM amplitudes for all investigated compounds.
The value obtained for $\bar{Q}$ for LuNiO$_3$ of $0.073$~\r{A} is in very good agreement with available experimental data ($Q_{exp}=0.075$~\r{A}~\cite{Alonso:2001bs}).
PrNiO$_3$ seems to be very close to the transition, as its $\bar{Q}$ value is very close to the MIT, and the stable BM would eventually be lost if a reduced $U$ is used for PrNiO$_3$, as suggested by our cRPA calculations~\cite{hampel_ederer:2018}.
Moreover, previous studies find that the magnetic order appears to be crucial in stabilizing the BD phase in \rno{Pr} and \rno{Nd} \cite{Medarde1997,Ruppen2017,Haule2017,hampel_ederer:2018}.
The stability and influence of the magnetic order goes beyond the scope of our work and
requires further investigation.
However, the overall trend of an increase in $Q$ and in the stability of the BM through the series for smaller $R$-cations
is consistent with experiments and in line with earlier studies~\cite{Varignon2017,hampel:2017,hampel_ederer:2018}.

%%%%%%%%%%%%%%%%%%%%%%%%%%%%%%%%%%%%%%%%%%%%%%%%%%%%%%%%%%%%%%%%%%%%%%%%%%%%%%%%
%
% Results: Landau theory
%
%%%%%%%%%%%%%%%%%%%%%%%%%%%%%%%%%%%%%%%%%%%%%%%%%%%%%%%%%%%%%%%%%%%%%%%%%%%%%%%%
\textit{Landau theory.}
We finally show that the main qualitative features of the MIT found above can be rationalized in terms of a
Landau theory, which involves two coupled scalar order parameters (OP):
the ED, $\nu$, and an additional OP, $\phi$, which distinguishes between metallic
(conventionally associated with $\phi>0$) and insulating behavior ($\phi<0$).
The reason why this second OP is required is clear from the results above: a non-zero value of
$\nu$ can correspond either to a metallic phase (at small values of the
on-site modulation, $\Delta_s$, or, equivalently, of $Q$), or to an insulating one.
In other words, a metallic monoclinic phase with a non zero-value of the BM amplitude $Q$ is in principle possible,
in agreement with recent experimental findings~\cite{middey_chakhalian_2018}.
Such an OP has been introduced to describe the Mott transition in the DMFT framework,
in analogy to the liquid-gas transition \cite{Kotliar2000,Limelette2003}.
Note that the present Landau theory aims at describing the MIT between the two PM phases,
while the earlier Landau descriptions \cite{Lee2011,*Lee2011b,Ruppen2017} aimed at the magnetic transition
(see also Ref.~\cite{han_millis_2018} in relation to ruthenates.)

Assuming  the simplest coupling allowed by symmetry, $\phi\nu^{2}$,
the PM transitions can be described by the following energy functional:
$F[\nu,\phi]\,=\,F_\nu + F_\phi + \lambda\phi\nu^2$, with
$F_\nu =  \frac{1}{2} b\nu^2 + \frac{1}{4}c\nu^4 - \Delta_s\nu + \frac{1}{2} \kappa \Delta_{s}^{2}$,
$F_\phi = \frac{1}{2} a\phi^2 + \frac{1}{4}u\phi^4 - h\phi$, where $u, c > 0$ and
the coupling parameter $\lambda > 0$ is of order one. The coupling to the lattice is
represented by the linear term $\Delta_s \nu$, with $\Delta_s$ serving
as a symmetry-breaking field
(alternatively, the BM can be introduced with $Q = g \Delta_s$).
In the absence of ED, the system is a metal, so that we must assume $a$ and $h$ to be positive.
$b$ is a key control parameter related to $\chi_{e}$, and it depends critically on external parameters
such as, e.g., the bandwidth, $b = b_0(W - W_{c})$.

\begin{figure}
   \includegraphics[width=1.0\linewidth]{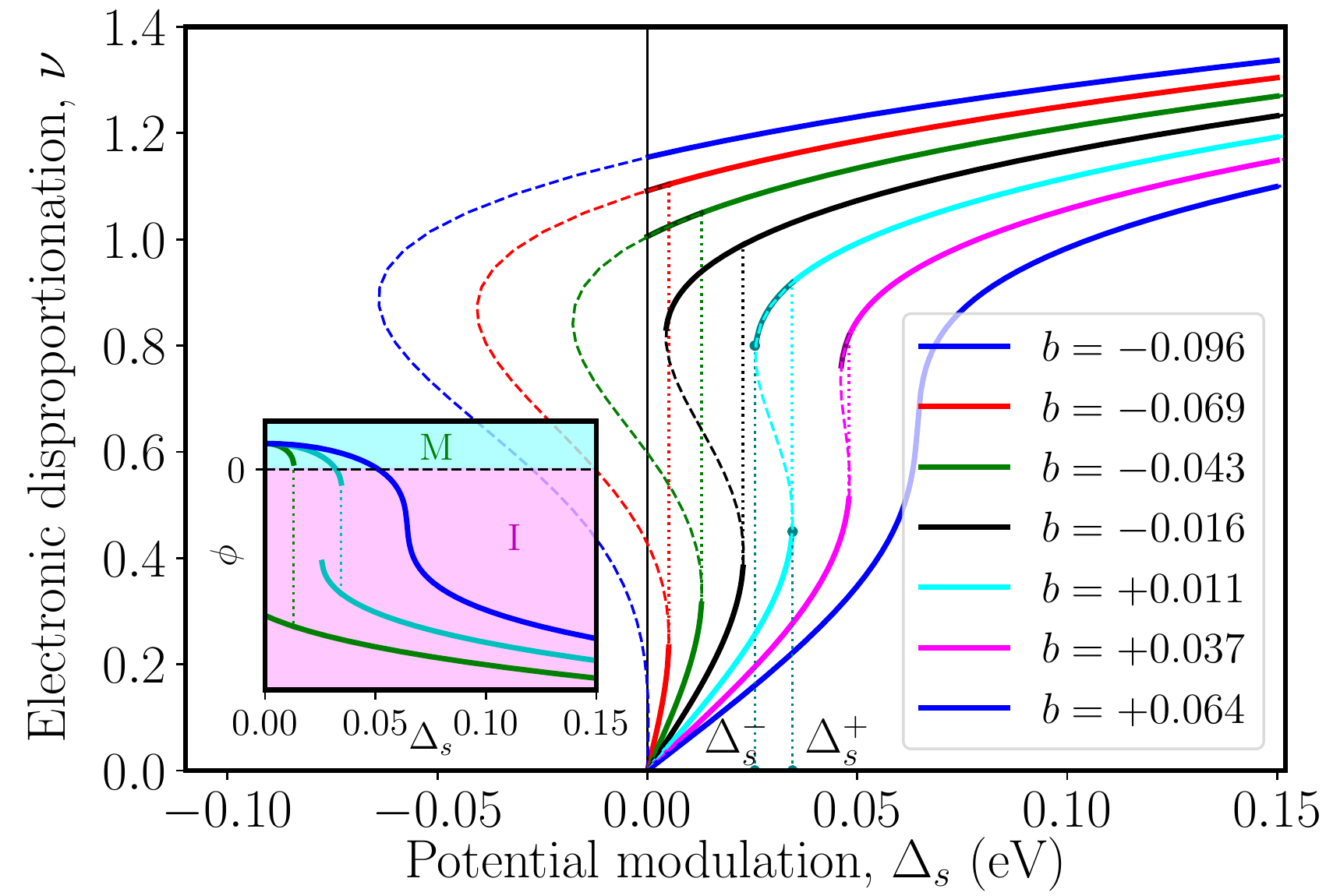}
   \caption{ED, $\nu$, as a function of $\Delta_{s}$ for various values of parameter $b$,
   as obtained within the Landau theory described in the text. Dashed lines display
   unstable solutions. Critical points $\Delta_{s}^{-}$, $\Delta_{s}^{+}$ are indicated
   for one of the cases. Inset: MI OP $\phi$ as a function of $\Delta_{s}$ for three
   selected cases ($b = -0.043$, $0.011$, $0.064$). Region marked with ``M'':
   metallic phase ($\phi > 0$); region marked with ``I'': insulating phase ($\phi < 0$).
   }
   \label{fig:LT_nu_vs_deltas}
\end{figure}

Without loss of generality we can set $u = c = a = 1$.
Minimizing $F[\nu,\phi]$ yields the following coupled equations of state:
\begin{align}
&b\nu + \nu^3 + 2\lambda\phi\nu = \Delta_s
\label{eq:eqstatenu}\\
&\phi + \phi^3 = h - \lambda\nu^2
\label{eq:eqstatephi}
\end{align}
The numerical solution of these equations is displayed in Fig.~\ref{fig:LT_nu_vs_deltas}, for varying $b$.
For $b > 0$, starting from a value $\phi = \phi_{0} > 0$, typical for the metallic phase at $\nu=0$, and
increasing $\Delta_s$ leads to a strongly non-linear dependence of $\nu$ on $\Delta_s$, with $\phi$ continuously decreasing
(because of the $-\lambda \nu^{2}$ term)
and gradually reaching negative values (inset of Fig.~\ref{fig:LT_nu_vs_deltas}).
At a critical value of $b$, the $\nu[\Delta_s]$ curves acquire a vertical tangent and beyond this value,
an S-shape with an unstable branch is found,
typical of a first-order transition, with two vertical tangents delimiting the
two spinodal values of $\Delta_s$, $\Delta_s^{-}$ and $\Delta_{s}^{+}$.
When $b$ is further decreased, a spontaneous instability is found, with a
jump of $\nu$ to a finite value for an infinitesimal $\Delta_s$.
This general behavior is in excellent qualitative agreement with Fig.~\ref{fig:bd_vs_q_model}.
A more detailed analysis (see \cite{suppl}) reveals that both spinodal values scale
as $\Delta_{s}^{*} \sim (b - b^{*})^{\frac{3}{2}} \sim (W - W^{*})^{\frac{3}{2}}$,
which is apparent in Fig.~\ref{fig:q_vs_w}.

%%%%%%%%%%%%%%%%%%%%%%%%%%%%%%%%%%%%%%%%%%%%%%%%%%%%%%%%%%%%%%%%%%%%%%%%%%%%%%%%
%
% Fig4a et b
%
%
%%%%%%%%%%%%%%%%%%%%%%%%%%%%%%%%%%%%%%%%%%%%%%%%%%%%%%%%%%%%%%%%%%%%%%%%%%%%%%%%

%%%%%%%%%%%%%%%%%%%%%%%%%%%%%%%%%%%%%%%%%%%%%%%%%%%%%%%%%%%%%%%%%%%%%%%%%%%%%%%%
%
% Section: Conclusions
%
%%%%%%%%%%%%%%%%%%%%%%%%%%%%%%%%%%%%%%%%%%%%%%%%%%%%%%%%%%%%%%%%%%%%%%%%%%%%%%%%
\textit{Conclusions.}
We have presented a theory of the combined structural and electronic MIT in
bulk \rno{$R$}.  The driving force is the proximity to the electronic
disproportionation instability, which is cooperatively reinforced by the
coupling to the lattice breathing mode. The transition is thus controlled both
by the electronic charge susceptibility and by the stiffness of this mode. The
key non-linearities associated with this cooperative effect can be rationalized
in terms of a Landau theory.  Our work provides a pathway to understanding the
MIT in other geometries, such as ultrathin films and heterostructures, and is
likely to have general applicability to other materials with a strong interplay
between electronic correlations and lattice degrees of freedom.

\acknowledgments{
We are grateful to C.~Ahn, D.~Basov, J.~Chakhalian, A.~Georgescu, A.~J.~Millis, J.~M.~Triscone, D.~van der Marel for discussions.
This work was supported by the European Research Council (ERC-319286-`QMAC'),
the Swiss National Science Foundation (NCCR MARVEL), the Swiss National Supercomputing Centre
(projects s575, s624, mr17, s820), and the Forschungsf\"orderungsgesellschaft (FFG)
COMET program IC-MPPE (project No 859480).
The Flatiron Institute is a division of the Simons Foundation.}

%
% Bibliography
\bibliography{refs}

\end{document}